%% file: 0-main.tex
\documentclass[acmtog]{acmart}

\usepackage{booktabs} 
\usepackage{multirow}
\usepackage[ruled]{algorithm2e} 

\SetAlFnt{\small}
\SetAlCapFnt{\small}
\SetAlCapNameFnt{\small}
\SetAlCapHSkip{0pt}

\graphicspath{{"./figures"}}

\setcopyright{acmcopyright}
\setcopyright{acmlicensed}
\setcopyright{rightsretained}
\setcopyright{usgov}
\setcopyright{usgovmixed}
\setcopyright{cagov}
\setcopyright{cagovmixed}

\input{macros}

\begin{document}
	\title{DJM: Compact Base Meshes for Displacement Mapping\\using Triangle Jacobians}
	
	\author{Congyi Zhang}
	\affiliation{\institution{University of Texas at Dallas}\country{United States}}
	\author{Nicholas Vining}
	\affiliation{\institution{NVIDIA}\country{Canada}}
	\affiliation{\institution{University of British Columbia}\country{Canada}}
	\author{Yanhong Lin}
	\affiliation{\institution{The University of Hong Kong}\country{Hong Kong}}
	\author{Alireza Khatami}
	\affiliation{\institution{University of Texas at Dallas}\country{United States}}
	\author{Ziyu Sun}
	\affiliation{\institution{University of British Columbia}\country{Canada}}
	\author{Xiaohu Guo}
	\affiliation{\institution{University of Texas at Dallas}\country{United States}}
	\author{Wenping Wang}
	\affiliation{\institution{Texas A\&M University}\country{United States}}
	\author{Alla Sheffer}
	\affiliation{\institution{University of British Columbia}\country{Canada}}
	
	\begin{teaserfigure}
		\begin{center}
			\includegraphics[width=0.9\linewidth]{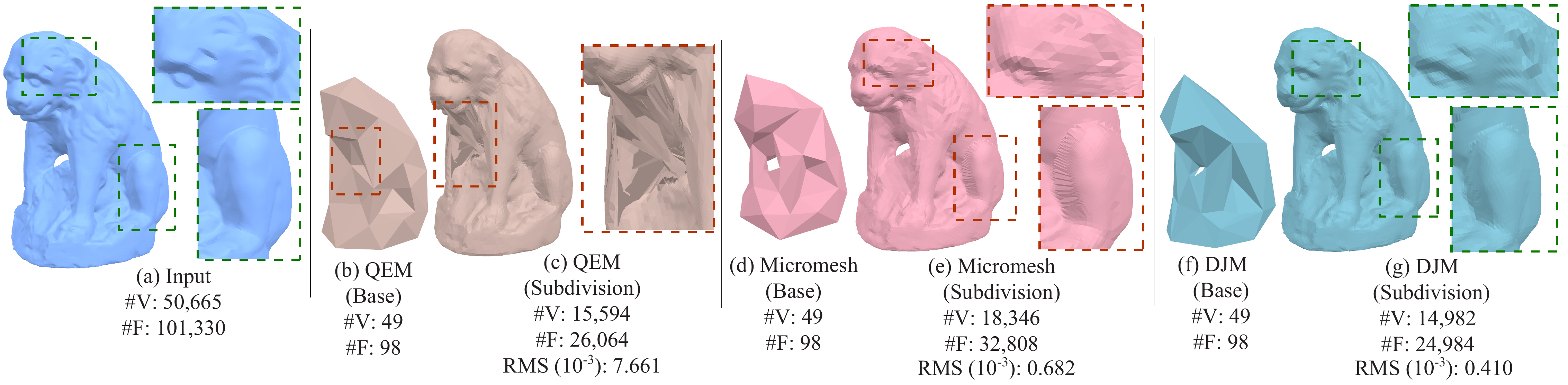}
		\end{center}
		\caption{We compactly represent input shapes (a) as a combination of a coarse base-mesh (d, left) and a displacement map enabling accurate reconstruction (d, right). Our DJM method outperforms reconstruction using standard QEM \cite{Garland1997QEM} simplification to construct the base (b), and the state of the art method of \cite{Maggiordomo2023MicroMesh} (c).  All methods use same size base-meshes. DJM uses fewer output faces than the alternatives yet achieves notably better reconstruction accuracy both visually and as measured by chamfer distance to the input.
		}
		\label{fig:teaser}
	\end{teaserfigure}

\input{1-abstract.tex}

	%
	%

\ccsdesc[500]{Computing methodologies~Mesh geometry models}
	
	%
	%
	
	\maketitle

\input{2-introduction}

	\input{3-related_works}

	\input{4a-problem_statement}

	\input{4b-base_mesh_construction}

\input{4c-inverse_problem}

\input{4d-subdiv_mesh_reconstruction}
	\input{5-results}

\input{6-conclusion}

	\input{7-acknowledgement}
	
	\bibliographystyle{ACM-Reference-Format}
	\bibliography{ref}
	
\end{document}

%% file: macros.tex
\usepackage[labelsep=period,skip=1pc,size=small]{caption}
\usepackage{subcaption}
\usepackage{booktabs}
\usepackage{enumitem}
\usepackage{cancel}
\usepackage{soul}

\newcommand{\point}{\mathbf{p}}
\newcommand{\qpoint}{\mathbf{q}}

\newcommand{\vertex}{\mathbf{v}}
\newcommand{\dir}{\mathbf{d}}
\newcommand{\disp}{h}
\newcommand{\tdisp}{t}
\newcommand{\mesh}{\mathcal{M}}
\newcommand{\vertices}{\mathcal{V}}
\newcommand{\faces}{\mathcal{F}}
\newcommand{\x}{\mathbf{x}}
\newcommand{\func}{\mathbf{f}}
\newcommand{\jac}{\mathbf{J}}

\usepackage{multicol}

\usepackage{longtable}  

\usepackage{lipsum}

%% file: 1-abstract.tex
\begin{abstract}
Representing complex geometry as a displacement function defined over a coarse base mesh enables compact storage and accelerated rendering. The core challenge in converting detailed triangle meshes into this representation is computing base meshes that have as few triangles as possible, while also supporting displacement functions that accurately approximate the input. Accurate approximation requires the supported displacement functions to bijectively map the input surface onto the base with low parametric distortion. We observe that this distortion can be measured by evaluating the pointwise Jacobian of the displacement functions. Our new DJM (Displacement Jacobian Metric)-based base-mesh construction method uses the Jacobian of the displacement functions to guide base mesh computation, enabling us to outperform prior approaches in terms of accuracy to size trade-off. We achieve this goal by proposing a variant of the QEM-based simplification scheme that constrains the displacement mapping between the input and the base to be bijective and low distortion (defined as satisfying a lower bound on the mapping Jacobian). When evaluating and encoding the displacement maps, we avoid unreliable ray-mesh intersections by explicitly storing the mapping between the input mesh and the base throughout the construction process, and use this mapping within a robust inverse barycentric displacement solver to obtain dense base-to-mesh correspondences to assist all computations. We demonstrate DJM to outperform alternative schemes in terms of reconstruction accuracy to size trade-off, and demonstrate its robustness and usability for micromesh based rendering and neural encoding.

\end{abstract}

%% file: 2-introduction.tex
\section{Introduction}
\label{sec:intro}
\begin{figure}
	\includegraphics[width=0.95\linewidth]{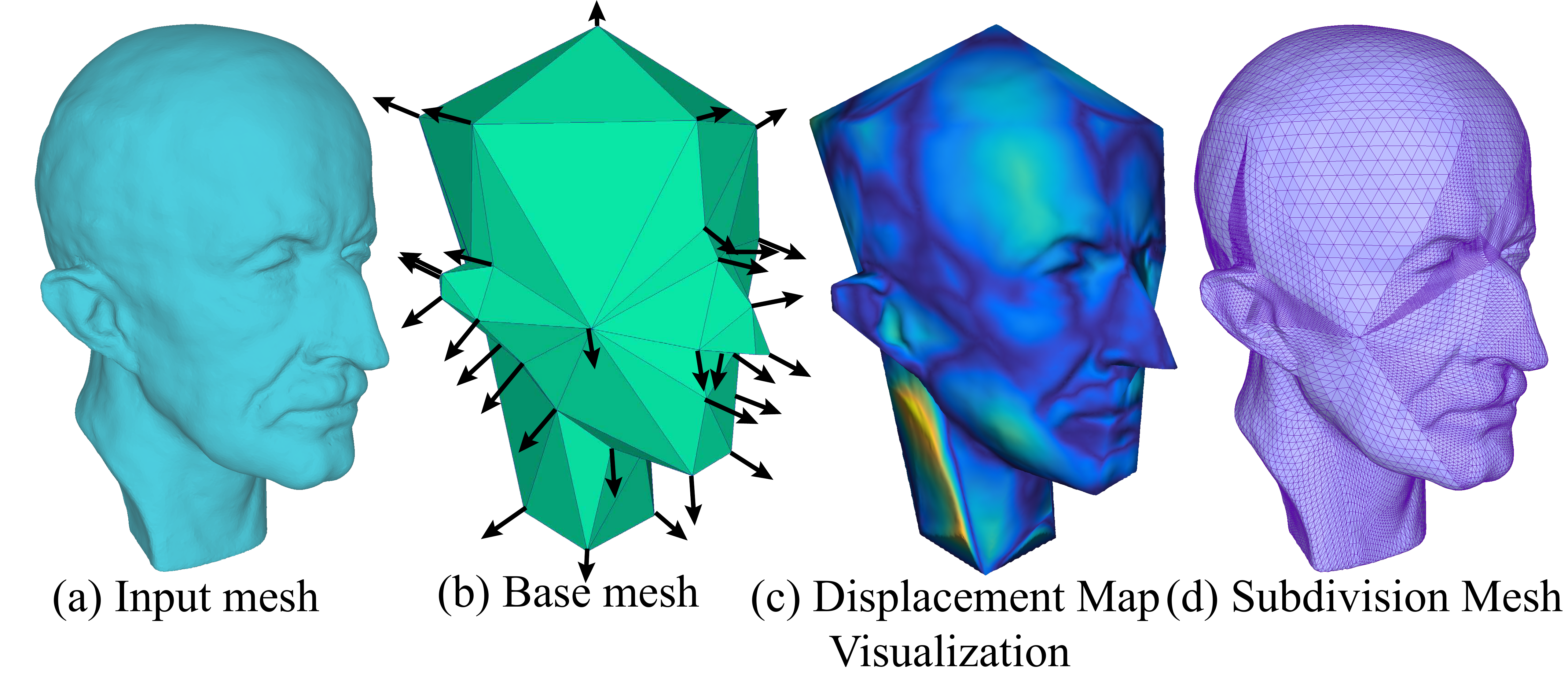}
\caption{Representing shapes (a) as a base mesh (b, arrows visualize corner displacement directions) and a displacement map (c, visualized as offset along displacement directions defined at base corners), facilitate fast subdivision based reconstruction and rendering (d). Representation computed using our DJM method.}  
\label{fig:displacement}
\end{figure}

Displacement mapping methods compactly represent complex shapes as a combination of a coarse {\em base mesh} and a displacement map (Fig~\ref{fig:displacement}); instead of storing the full geometry only the base and the displacement are stored. The displacement map is a function on the surface of the coarse base mesh that encodes the high-resolution geometric detail of the input in the form of an offset along a computed direction \cite{nvidia2022ada,cook1984shade}.  
Scalar displacement mapping \cite{cook1984shade}  uses base mesh face's normals as the displacement directions \cite{cook1984shade}.  Micromesh representations \cite{nvidia2022ada, Maggiordomo2023MicroMesh}  compute these directions by interpolating direction vectors defined at base mesh vertices (Fig~\ref{fig:displacement}b).
 Displacement-based representations
are useful for a wide range of applications in computer graphics, including geometry compression \cite{Pentapati2025MeshCompression,Dou2024DiffMicroMesh} and displacement mapping for rendering \cite{nvidia2022ada,Thonat2021SIGA,Thonat2023SIGA}. The usability of a representation depends on two factors - the accuracy at which it captures the input shape and the size of the resulting shape encoding. We present a novel displacement based representation that achieves significantly better accuracy-to-size trade-off than state of the art alternatives (Fig~\ref{fig:teaser}).

The key factors that impact representation accuracy are mapping bijectivity and distortion. If a mapping is not bijective, the reconstruction will miss parts of the input surface, and if the mapping has high distortion, surface reconstruction will be lossy \cite{Sander}. 
These requirements are easier to satisfy for fine base meshes, which remain close to the input surface, and get harder for coarser base meshes. Early methods for displacement mapping had therefore utilized very fine meshes (Sec~\ref{sec:related}). Several recent methods \cite{Pentapati2025MeshCompression,Dou2024DiffMicroMesh} use the QEM simplification method \cite{Garland1997QEM} to generate coarse base meshes and use a range of strategies to optimize the mapping relative to the base.  Maggiordomo et al \shortcite{Maggiordomo2023MicroMesh} propose a base mesh simplification method specifically targeting displacement mapping as an application. Neither QEM nor \cite{Maggiordomo2023MicroMesh} base construction approaches explicitly optimize for bijectivity or low distortion, leading to significant reconstruction artefacts at coarser base resolutions (Fig~\ref{fig:teaser}bc). Our {\em Displacement Jacobian Metric} (DJM) based method simultaneously computes base-meshes and corresponding bijecitve, low distortion displacement mappings, enabling superior reconstruction (Fig~\ref{fig:teaser}d). 

Our method uses the QEM edge-collapse based simplifier as a starting point, but modifies it to achieve bijectivity and minimize distortion. Like other recent methods, we use {\em direction vectors} defined at base mesh vertices to define the displacement mapping, interpolating those barycentrically across each base face. We observe that given a set of triangles expected to map to a base mesh face and the direction vectors at the base-mesh corners we can directly evaluate mapping bijectivity and explicitly and effectively compute a Jacobian-based metric that directly measures the displacement distortion. We use this metric to bound the distortion introduced by our simplifer, and disallow edge collapses that would introduce unacceptable distortion. We similarly disallow collapses that violate bijectivity. Supporting these evaluations throughout the simplification process requires explicit tracking of the correspondence between the original surface and the base mesh. Traditionally, base-to-input displacement mapping is computed via ray-casting, or shooting rays from the base along the displacement direction and computing their intersections with the input surface. The further the base deviates from the input the more likely the computation to fail (Sec~\ref{sec:mapping}). We avoid the need for ray-casting throughout all stages of our method, replacing them with a combination of forward and backward mapping solvers (Sec~\ref{sec:mapping}. Our backward mapping finds for each given vertex $p$ on the input surface, its corresponding base triangle on the base mesh and a set of non-negative barycentric coordinates $\{ \alpha,\beta, \gamma \}$ and a displacement $h$ sufficient to reconstruct $p$. Our forward solver uses the computed backward mapping to robustly locate for each point on the base mesh the corresponding point on the input. These computations ensure that each base mesh face is continually associated with an input mesh region, supporting bijectivity and distortion evaluations, as well as the final computation of surface to base correspondences used to encode the displacements across base faces.   

We validate DJM by applying it to 109 diversely sourced high resolution  inputs, and use it to generate subdivision meshes  well-suited for fast rendering (Figs~\ref{fig:teaser}d, \ref{fig:displacement}d) and neural encodings (Sec~\ref{sec:neural_results}).
We compare micromeshes generated using DJM to those generated using QEM and \cite{Maggiordomo2023MicroMesh}. Our outputs more accurately approximate the inputs using smaller size meshes. We similarly compare DJM-based neural encodings to outputs of base-mesh based state of the art methods of \cite{Sivaram2024NGF,Pentapati2025MeshCompression} demonstrating similarly superior accuracy-size tradeof. 
 Our code and inputs are available at \url{https://www.cs.ubc.ca/labs/imager/tr/2026/djm/}.

%% file: 3-related_works.tex
\section{Related Work}
\label{sec:related}

\paragraph{Geometry Compression.}
Mesh compression is important for scenarios where large geometry must be stored in memory or disk, or transmitted over a network, \cite{alliez2005recent, maglo20153d}. The outputs of these methods must be decompressed before processing. In contrast displacement based representations are designed to be processed in their {\em compressed} form.  

\paragraph{Mesh Simplification.} Mesh simplification methods aim to produce lower-resolution meshes  that capture the geometric, topological, or semantic properties of the original shape \cite{Hoppe1996PM,hoppe1993mesh,Garland1997QEM}. Typically, this simplification is computed by a repeated series of edge collapses \cite{Hoppe1996PM,hoppe1993mesh} ordered so as to minimize a per collapse error metric. The most commonly used quadric error metric (QEM) \cite{Garland1997QEM} measures the deviation of a post-collapse vertex from a set of planes it is associated with. 

While one can use QEM or other simplification methods as-is to compute a displacement base mesh, this poses two challenges.   
First, all traditional simplification methods  attempt to minimize a geometric error between the original and simplified meshes, not between the original mesh and the displaced mesh built on top of the simplified one; these two errors are not identical. Second and more critically, without an explicit bijective mapping from the simplified to the original mesh, existing  edge-collapse based methods, such as QEM or progressive meshes \cite{Hoppe1996PM} only accumulate information from the original endpoints into the new collapsed vertex. This leads to a gradually biased metric, since the current edge may not be accurately represented by the accumulated information, especially when the mesh is extremely simplified. In contrast, our method tracks this correspondences throughout the simplification process, which gives much more accurate bijectivity testing and distortion control, see e.g Fig~\ref{fig:teaser}b versus d. 
\paragraph{Base Mesh Parameterization and Correspondence Tracking.} A large body of work computes mappings between 3D shapes and base meshes, and uses different strategies to compute the base \cite{Kraevoy2004CrossParameterization,Lee1998MAPS}. These methods store the 3D coordinates of the mapped vertices over the base mesh, requiring roughly three times more storage space than displacement mapping, which stores a single offset. 
Prior research noted that  a bijective correspondence can be computed during mesh simplification by starting with the identity correspondence and updating it during each local edge operation via careful bookkeeping \cite{Lee1998MAPS,liu2020subdivision,Liu2023Simplification,Sharp2019Navigating}, a strategy which we also employ. Liu et al. ~\shortcite{Liu2023Simplification} build a base mesh for geometric multigrid problems by simultaneously computing both a base mesh and a parameterization of the input mesh to its simplified counterpart (a process introduced in a previous work \cite{liu2020subdivision} commonly called "successive self-parameterization"). They use intrinsic surface triangulation as a base mesh and each face is essentially a patch of the input surface, rather than a flat polygon. This approach is not suitable for displacement based representations as linearly interpolating displacement vectors (and vertex coordinates) across intrinsic triangles produces invalid results.
Nanite \cite{Karis2021Nanite} generates a correspondence in the form of a cluster hierarchy that can be used for screen-space geometry level of detail; the aim of this cluster hierarchy is to produce crack-free outputs for rendering from a given camera position, and not manifold geometry or globally accurate reconstructions. When constructing the cluster hierarchy they use QEM to select edges to collapse.
Zhang et al. ~\shortcite{zhang2023progressive} produce base meshes with correspondences, but optimize for very different objectives than our own, specifically rest-shape recovery and intrinsic prolongation for FEM simulation.

NGF \cite{Sivaram2024NGF} use a quad base mesh computed by first applying QEM and then pairing triangles to form quads, they then learn a correspondence between the base and the 3D surface using differentiable rendering. Their representation encodes the input surface as 2D to 3D functions over the base faces. Our method supports neural encoding and only stores an offset relative to the base instead of full 3D coordinates, allowing better accuracy to size trade-of (Fig~\ref{fig:NGF_comparison}, Sec~\ref{sec:results}). Yang et. al \shortcite{yang2025neupps} construct coarse quadrilateral base meshes and encode the 3D surfaces using a 2D to 3D parameterization.   
Displacement mapping allows for more compact storage then general parameterization based approaches as it stores one value per base mesh point instead of three.

\paragraph{Displacement Mapping and Displaced Surfaces.} Displacement mapping \cite{cook1984shade} augments low-resolution base meshes with fine geometric details encoded via scalar or vector displacement functions. Early use of 
displacement mapping targeted adding fine details to surfaces, storing these details as texture maps \cite{Lee2000DisplacedSubdivisionSurfaces}. Historically, real-time rendering has focused on scalar displacement mapping, as it requires only a single channel and can be easily filtered \cite{donnelly2005per,hirche2004hardware}; scalar based displacement mappings require fine base meshes as larger deviation between the base and the input leads to visible cracks along seam boundaries and deviations between the base surface normal and the normal of the displaced surface \cite{Niessner2013Displacement}. Modern computer graphics hardware enables algorithms for rendering displacement mapped surfaces directly on the GPU.

Storing geometry as a base representation and displacement maps for compression purposes dates back to Krishnamurthy and Levoy ~\shortcite{krishnamurthy1996fitting} who store data as vector-valued displacements over a B-spline patch network, and Lee et al. ~\shortcite{Lee2000DisplacedSubdivisionSurfaces} who represent detailed surface models as a combination of a base mesh and a scalar-valued displacement function stored over it. The modern evolution of this approach is NVIDIA's micromeshes \cite{nvidia2022ada} which address the problems of previous real-time displacement mapping methods by providing a geometry representation that is efficiently compressible, has no seams, and can be rendered from its compressed form in real-time. They store displacements as a combination of a per-vertex displacement direction and a per-triangle height function stored in barycentric coordinates \cite{Yuksel2010MeshColors}; the displaced location of a point on a triangle is given by barycentrically interpolating the displacement directions, then offsetting in that direction by the sampled height. While NVIDIA's white paper describes the micromesh format and its rendering, it does not establish how to generate base meshes and displacement maps suitable for micromesh rendering.   

\begin{figure}[h]
	\centering
	\includegraphics[width=0.95\linewidth]{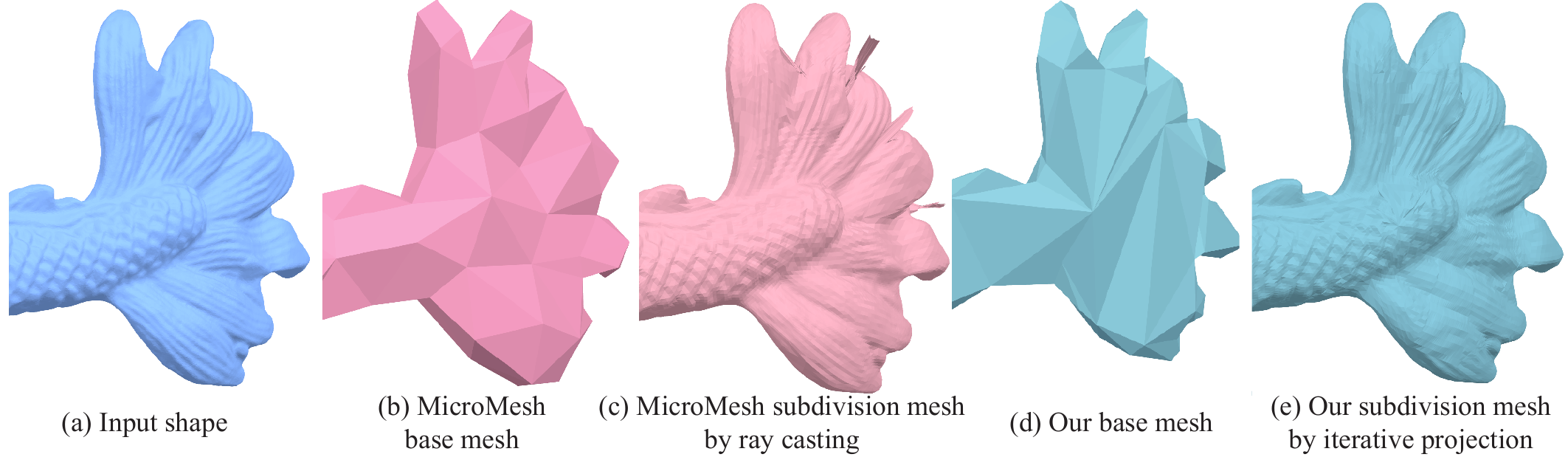}
	\caption{Ray casting-based methods are not stable for displacement calculation.}
	\label{fig:ray_casting}
\end{figure}

Maggiordomo et al. \shortcite{Maggiordomo2023MicroMesh} propose an edge-collapse based simplification process for micromesh base-mesh construction. They use a metric that balances triangle quality, quadric error, normal deviation and approximate bijectivity.  Their method does not maintain a correspondence between the input and base mesh, and displacements are calculated by ray-casting after the base mesh is constructed.  Consequently the method is susceptible to catastrophic errors when the ray-tracing fails to intersect the surface at the desired location (Fig.~\ref{fig:ray_casting}). DJM avoids such failures by maintaining a consistent input to base mesh correspondence, avoiding ray-casting. It consistently outperforms Maggiordomo's method in terms of accuracy to size ratio (Fig.~\ref{fig:teaser}, Sec.~\ref{sec:results}). 
Dou et al. ~\shortcite{Dou2024DiffMicroMesh} use a differentiable renderer to optimize an existing base mesh for micromesh construction and use QEM to compute the base. Their method is complementary to ours and can be applied as a DJM post-process (as no code is provided we cannot conduct this experiment).
\paragraph{Neural Displacement Maps.} Pentapati et al \shortcite{Pentapati2025MeshCompression} encode a neural displacement function on top of a coarse base mesh. They note that methods like NGF \cite{Sivaram2024NGF} that optimize for rendering loss do not guarantee geometric fidelity as the two are not the same. 
They address this discrepancy by using a bijective correspondence created by successive self-parameterization, and store the sampled values using a simple MLP. Their base meshes are generated using QEM as-is.  We demonstrate the suitability of our base meshes for neural displacement mapping in Sec. ~\ref{sec:results}.

%% file: 4a-problem_statement.tex
\section{Method}
\label{sec:method}
\subsection{Problem Statement}

Given a high resolution input triangle mesh $\mesh_I=\{V_I,F_I\}$, our goal is to compose a low resolution base mesh $\mesh_B=\{\tilde{V_B},\tilde{F_B}\}$, such that $\mesh_I$ can be represented as a displacement map defined on the base mesh $\mesh_B$. A normalized displacement direction $\dir$ is defined at each of the base mesh vertices $\tilde{V_B}$; the displaced position at each point  $\qpoint$ on a base mesh face $f$ with vertices $\vertex_1,\vertex_2,\vertex_3$  
is defined as 
\begin{equation}\label{eq:disp_map}
		\point=\mathbf{h}(\qpoint) = \alpha \vertex_1+\beta \vertex_2+\gamma \vertex_3+h\dir,
\end{equation}

where $\{\alpha,\beta,\gamma\}$ are the barycentric coordinates of  $\qpoint$,  $h$ is the displacement value at  $\qpoint$, and $\dir$ is the normalized vector  $\frac{\alpha\dir_1+\beta\dir_2+\gamma\dir_3}{\|\alpha\dir_1+\beta\dir_2+\gamma\dir_3\|}$. 

Our challenge is to compute the base mesh and vertex directions such that the displacement mapping  $\mathbf{h}$ from $\mesh_B$ to $\mesh_I$ is (1) bijective and (2) has bounded distortion, and (3) that the base mesh triangles are sufficiently well shaped. Most importantly we want the size, or face count, of the base mesh $\mesh_B$ to be small, and as close as possible to some target size $S$ while still satisfying conditions (1) through (3). 

Our method for constructing the base mesh conceptually follows the same progressive simplification framework as QEM or Maggiordomo et al. ~\shortcite{Maggiordomo2023MicroMesh}, but with a different optimization schedule and constraints that jointly ensure conditions (1) through (3) are strictly satisfied (Sec~\ref{sec:base}). Our computations throughout this process, and when using the resulting displacement mapping, are based on an explicit point-to-point correspondence between the base and input meshes which we compute as described in Sec~\ref{sec:tracking}.  Throughout the process we explicitly avoid performing any ray-casting operations, used extensively by prior methods. Ray casting is problematic for two reasons. First, it is unstable when the ray is tangential or exactly through an edge of the target mesh triangle; second, it is also ambiguous: we do not {\em a priori} know whether the input mesh is in front or behind a base triangle, and thus we have no principled way of discerning between in front and behind intersections (distance to intersection can be an unreliable predictor).  
\paragraph{Displacement Mapping Application}
Once the construction of the base mesh is complete, we can use the displacement mapping to store and apply micromesh displacements \cite{nvidiamicromeshingwhitepaper} or learn a neural encoding. For micromesh displacement mapping, we find a maximum subdivision level and sample locations on the base triangles of $\mesh_B$ using the same strategy as Maggiordomo et al. \shortcite{Maggiordomo2023MicroMesh}; we then compute displacements for each sample location using our displacement mapping as discussed in Sec~\ref{sec:forward}. For neural encoding (Sec~\ref{sec:neural_results}) we similarly need to sample points on the base and compute their displacements, using the same process.

%% file: 4b-base_mesh_construction.tex
\subsection{Base Mesh Construction}
\label{sec:base}
\begin{figure}[h]
	\centering
	\includegraphics[width=0.8\linewidth]{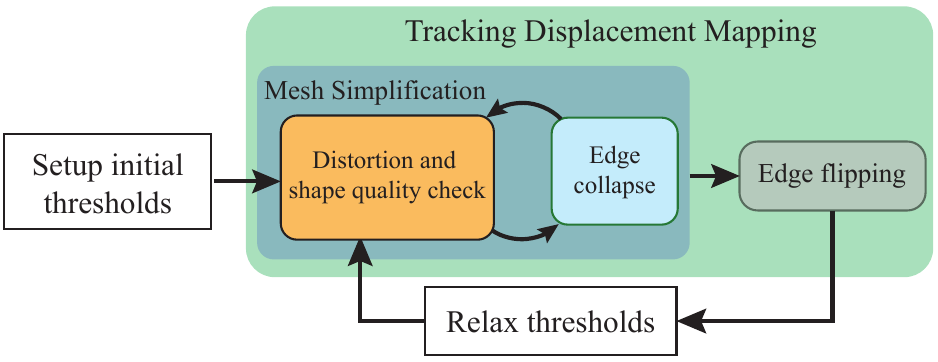}
	\caption{DJM base mesh construction workflow.}
	\label{fig:workflow}
\end{figure}

Our base mesh construction method loosely follows the standard QEM workflow \cite{Garland1997QEM}, but seeks to balance the classical QEM geometric collapse metric versus mapping distortion and triangle shape. 
One naive approach to address these additional desiderata is to run QEM as-is, but reject any collapse operation that introduces base mesh triangles that violate a tight distortion or triangle shape threshold. In practice, this approach introduces many base mesh faces with distortion or shape metrics just barely above the threshold. Another alternative would be to modify the collapse metric to combine geometric fidelity with distortion and/or shape metric terms; the challenge here would be to meaningfully balance the different terms as they measure hard to compare properties. 

We avoid the drawbacks of both approaches by adopting a progressive relaxation strategy. We start the process with strict distortion and shape thresholds and use the standard QEM metric to prioritize edges for collapse, preventing collapses that violate either threshold. Once no more edges can be collapsed with these thresholds in place we relax the thresholds slightly, and continue the collapse process. We repeat the relaxation step until the mesh reaches the desired size or until any further collapses would violate one of our hard distortion and shape thresholds (whichever happens first) (Fig~\ref{fig:workflow}).
In addition to incorporating distortion and shape quality metrics into the simplification framework, we explicitly prevent edge collapses that prevent the displacement function  from being bijective. 

As we seek to minimize distortion and optimize triangle shape at the end of each edge collapse iteration and before relaxing the thresholds, we perform a series of edge flip operations if these flips do not violate mapping bijectivity and improve distortion without significantly degrading shape, or vice versa (Sec~\ref{sec:flip}). 
\paragraph{Tracking the Displacement Mapping} The computations above are made feasible by a self-parameterization which tracks the mapping between the input and base meshes throughout the simplification process. The mapping is initialized as the identity mapping when the process starts, and is updated during every subsequent collapse operation (Sec.~\ref{sec:tracking}). The process operates on the assumption that all vertices mapped to the umbrella of the edge being collapsed can be mapped to the 1-ring neighbourhood of the new vertex that replaces this edge using the displacement map. If this mapping is invalid or non-bijective, the collapse is rejected. 

The key metric that supports our distortion evaluation is the displacement Jacobian, computed as discussed in Sec~\ref{sec:jac}. We use the longest edge to the corresponding height ratio as a metric of triangle shape. 
\paragraph{Computing Per-Vertex Displacement Directions} 
When considering a candidate edge for collapse, we compute the direction for the vertex replacing it using a normal cone \cite{Maggiordomo2023MicroMesh}. Unlike Maggiordomo et al., we precisely know the region on the input mesh that maps to the 1-ring neighbourhood of the edge being collapsed; therefore, we can compute the cone direction more accurately. We include the normals of all the faces in this region in the cone and use the cone's central vector as the vertex direction. This choice minimizes the worst angle between the base vertex and input mesh face normals. We check the angles between this computed displacement direction at the potential new vertex and the normals of the base mesh faces in its 1-ring neighbourhood.  If any of these angles becomes large, the displacement map can become highly distorted or non-bijective. We similarly check the difference between the normals of these base faces and the displacement directions of their other corner vertices. We reject the collapse if any of these angles is above $45^\circ$, with the threshold gradually relaxed to $80^\circ$.
\subsubsection{Jacobian-based Displacement Distortion Measurement}
\label{sec:jac}
We observe that the displacement map function \autoref{eq:disp_map} defines a mapping from the point $\qpoint=\alpha \vertex_1+\beta \vertex_2+\gamma \vertex_3$ on the surface to an arbitrary point $\point$ in Euclidean space; in order to measure distortion effectively, we need a closed-form solution for its Jacobian. We can rewrite this mapping as 
\begin{align}
	\point &= \qpoint + t \left( \alpha \dir_1 + \beta \dir_2 + \gamma \dir_3 \right) \\
	&= \qpoint + t \left( \dir_3 + \alpha (\dir_1 - \dir_3) + \beta (\dir_2 - \dir_3) \right)\\
	&=\qpoint + t \left(d_3 + A
	\begin{bmatrix}
		\alpha \\ \beta
	\end{bmatrix}\right),
	\quad\text{where}\quad A=\begin{bmatrix} \dir_1 - \dir_3 & \dir_2 - \dir_3 \end{bmatrix}
\end{align}
And because
\begin{equation}
	\qpoint=\begin{bmatrix}
		\vertex_1 & \vertex_2 & \vertex_3
	\end{bmatrix}
	\begin{bmatrix}
		\alpha \\ \beta \\ \gamma
	\end{bmatrix}
	=\vertex_3 + 
	\begin{bmatrix}
		\vertex_1 - \vertex_3 & \vertex_2 - \vertex_3
	\end{bmatrix}
	\begin{bmatrix}
		\alpha \\ \beta
	\end{bmatrix}
\end{equation}
We have
\begin{equation}
\begin{bmatrix}
	\alpha \\ \beta
\end{bmatrix}
= (B^\top B)^{-1} B^\top (\qpoint - \vertex_3),
\quad \text{where} \quad
B = \begin{bmatrix} \vertex_1 - \vertex_3 & \vertex_2 - \vertex_3 \end{bmatrix}
\end{equation}
Finally the mapping is defined as
\begin{equation}
\point(\qpoint) = \qpoint + t \left( \dir_3 + 
A
\left( (B^\top B)^{-1} B^\top (\qpoint - \vertex_3) \right)
\right)
\end{equation}
By direct calculation, the Jacobian matrix $\jac(\qpoint)$ is therefore:
\begin{equation}
\jac(\qpoint) = I + t A (B^\top B)^{-1}B^\top
\end{equation}
This is a closed-form expression with respects to the base triangle vertices $\{\vertex_1,\vertex_2,\vertex_3\}$, the vertex directions $\{\dir_1,\dir_2,\dir_3\}$ and the displacement $t$ with regard to the unnormalized interpolated displacement direction; to compute the distortion, we can simply compute the determinant $Det(\jac)$ and measure how close it is to 1.

\section{Tracking Displacement Mapping}
\label{sec:tracking}
We maintain correspondences between the base mesh $\mesh_B$ and the input mesh $\mesh_I$ throughout the simplification process. 
Since the correspondences need to be computed for every potential edge collapse operation, the computation needs to be robust and extremely efficient. Directly computing the correspondence from the base to the input via ray-casting (as done by Maggiordomo et al. ~\shortcite{Maggiordomo2023MicroMesh}) is both computationally expensive and not robust, failing in cases where rays are tangential or nearly-tangential to the input surface.

We avoid ray casting and generate our correspondence by directly computing the inverse mapping from the input to the base mesh. As Liu et al. ~\shortcite{Liu2021Multigrid} and other works also observe, we recall that for every edge collapse step we only need to update the correspondences in the 1-ring neighbourhood of the edge being considered for collapse; all vertex positions and directions along and outside the boundary of this 1-ring neighborhood remain unchanged.

Given the correspondences before a potential edge collapse, we find all impacted vertices $\vertices$ on the input mesh $\mesh_I$ using the 1-ring neighbourhood before the collapse. We need to map these vertices to the new 1-ring neighbourhood $\faces$ of the vertex replacing the edge. For a vertex $\point$ in $\vertices$ we do not {\em a priori} know which of the faces on the 1-ring neighbourhood $\faces$ it best maps to. We therefore compute the local barycentric coordinates for the vertex $\point$ with respect to each face in the neighbourhood using the formulation in Sec~\ref{sec:inverse_local}, and check which face it is most likely to be on. If the local barycentric coordinates are outside the range $[0, 1]$, it means that the "projected" point $\qpoint=\alpha\vertex_1+\beta\vertex_2+\gamma \vertex_3$ is outside the face. 

When $\qpoint$ falls on, or near, a face edge, the inside-outside test may become unreliable due to numerical accuracy. When this situation occurs, following \cite{Maggiordomo2024InverseBaricentric} we clamp the barycentric coordinates to $[0,1]$ and reconstruct the point using \autoref{eq:disp_map}. We recall that for any point $\point$ that can be projected inside a base mesh triangle, we expect a very low reconstruction error (measured by calculating the distance between the original point $\point$ and its  location computed via displacement form the base $\qpoint + \disp\dir$). If this reconstruction error is larger than some epsilon (in our case, $10^{-4}$),  the point must lie outside the base mesh triangle. 

For a point deemed inside a base mesh face, we compare the normalized interpolated direction $\dir=\frac{\alpha\dir_1+\beta\dir_2+\gamma\dir_3}{\|\alpha\dir_1+\beta\dir_2+\gamma\dir_3\|}$ with the normals of all input mesh faces in the 1-ring of $p$. If the angle between the direction and the normal is above $90^\circ$ this means that mapping the vertex to the face will result in a non-bijective mapping.  If this happens we reject the possibility of the vertex mapping to this face. 

After checking both conditions, there may still be multiple feasible base faces for a vertex $\point$ (e.g. when a point is on the edges of mesh). In such a case, we assign the vertex to the face with smallest displacement distance $\disp$ to $\point$.
\subsubsection{Edge Flipping}
\label{sec:flip}
Our edge-flipping pass operates on a priority queue of faces, ordered by worst-to-best Jacobian determinant of their two attached faces.
For each edge, we check if flipping it will improve the worst Jacobian determinant of its faces, or if it can reduce the highest vertex valence among impacted vertices without notably impacting the Jacobian determinant ($0.9<det(\jac)<1/0.9$). If either the local displacement map or valences can be improved, we accept the flip; otherwise, we reject it. We use valence as a proxy for triangle shape as we found it to be more predictive of overall triangle shape at the end of the flipping pass than direct shape assessment, consistent with suggestions in meshing literature \cite{surazhsky2003explicit}.  We use the same process for updating the mapping after a flip as we do for edge collapse (Sec~\ref{sec:tracking}), this time only considering the two old and two new base mesh faces. We similarly reject the flip if it leads to distorted or non-bijective mapping.

%% file: 4c-inverse_problem.tex
\subsection{Mapping Computation}
\label{sec:mapping}
Throughout our algorithm, we require a method to compute, for an input mesh vertex $\point$, its barycentric coordinates with respect to different base mesh faces, and the inverse mapping from base mesh faces to the input mesh. Sec. ~\ref{sec:inverse_local} addresses the first scenario;  Sec~\ref{sec:forward} discusses the computation of the mapping from the base to the input mesh. 
\subsubsection{Inverse Barycentric Displacement}
\label{sec:inverse_local}
Assuming that the base mesh vertices $\{\vertex_1,\vertex_2,\vertex_3\}$, displacement directions $\{\dir_1,\dir_2,\dir_3\}$, and input mesh point $\point$ are given, we must find valid $\{\alpha,\beta,\gamma,\disp\}$ that satisfy \autoref{eq:disp_map}. We assume the vertex directions are on the same side of the triangle normal so that interpolating them cannot create a zero-length vector (this condition is explicitly accounted for in our direction computation during edge collapse and collapses that violate it are disallowed). Notice that normalizing the interpolated direction complicates the problem; without loss of generality, we simplify the formulation by instead introducing a displacement $\tdisp=\frac{\disp}{\|\alpha\dir_1+\beta\dir_2+\gamma\dir_3\|}$ for an unnormalized direction, so that
\begin{equation}\label{eq:disp_map_unnorm}
	p=\alpha \vertex_1+\beta \vertex_2+\gamma \vertex_3+\tdisp (\alpha \dir_1+\beta \dir_2+\gamma \dir_3)
\end{equation}
As discussed in \cite{Maggiordomo2024InverseBaricentric}, the inverse barycentric displacement problem can be formulated and solved as the solution of a cubic equation. This closed-form solution provides a theoretical analysis of the problem based on the classification of three roots of the cubic equation. In practice, however, finding the real root is highly unstable due to numerical issues. Even with an advanced cubic root solver implementation that uses bisection on bounding intervals for roots \cite{Schneider2002GT,eberly_geometrictools}, the real root cannot be accurately and robustly computed. Additionally, as the general displacement map is not always bijective, we only require one root in practice (the solution $\tdisp$ with the smallest magnitude $|\tdisp|$ \cite{Maggiordomo2024InverseBaricentric} to ensure continuity), which means all other solutions are redundant; thus it is not really necessary to solve for all three roots analytically.

We consequently use a more efficient and stable iterative method to solve the inverse barycentric displacement problem. We firstly replace $\gamma$ as $1-\alpha-\beta$ and we have:
\begin{equation}\label{eq:func_x}
	 \left((\dir_1-\dir_3)\alpha+(\dir_2-\dir_3)\beta+\dir_3\right)\tdisp + (\vertex_1-\vertex_3)\alpha+(\vertex_2 - \vertex_3)\beta + \vertex_3 - \point = \mathbf{0}
\end{equation}
Then by denoting a variable $\x=(\alpha,\beta,\tdisp)^T$, \autoref{eq:func_x} defines a quadratic function $\func(\x)=\mathbf{0}$. Instead of directly finding a closed-form solution, we can now solve a non-linear least square problem $\|\func(\x)\|^2$ using the Gauss-Newton algorithm. By computing the Jacobian matrix $\jac$ of $\func$, we can iteratively approach the solution by
\[
\x^{(i+1)}=\x^{(i)} - \jac^{-1}\func(\x^{(i)})
\] 
We compute $\jac^{-1}\func(\x^{(i)})$ by using LU decomposition that solves the linear problem $\jac\Delta\x^{(i)}=-\func(\x^{(i)})$ in a least squares manner; when the matrix is close to non-invertible, we fallback to QR decomposition for a more robust solution.

To obtain a valid solution it is crucial to find a meaningful initial guess $\x^{(0)}$; we do this by setting $\tdisp=0$ and then solving for the projected barycentric coordinates $\alpha$ and $\beta$ of $\point$ on the plane of triangle $\vertex_1,\vertex_2,\vertex_3$. These initial barycentric coordinates may be outside the triangle since we project $\point$ along the triangle normal.
To ensure fast and robust convergence, we clamp the barycentric coordinates to $[0,1]$ and normalize them to ensure the initialization is inside the triangle. We terminate the iterative process once the residual is sufficiently low or a maximal number of iteration is reached. 
Finally, we compute $\disp=\|\alpha\dir_1+\beta\dir_2+\gamma\dir_3\|\tdisp$, which is the displacement value that measures the distance from the point $\point$ to the triangle along the interpolated direction.

We note that a similar iterative numerical strategy is also mentioned in \cite{Thonat2023SIGA}, however our formulation is different: instead of using their cross-product equation, we have found that directly solving the problem in this manner yields faster and more robust convergence.

%% file: 4d-subdiv_mesh_reconstruction.tex
\subsubsection{Base to Input Mapping}
\label{sec:forward}
Encoding the displacement mapping whether for micromesh subdivision or neural encoding requires locating for points $p$ on the base mesh the corresponding points on the input mesh and recording the displacement offset. 
As before, we  perform this task while avoiding ray-casting.

We sidestep ray-casting by exploiting our previously computed mapping from the input mesh $\mesh_I$ to the base mesh $\mesh_B$ and using an iterative closest point method. Specifically, we directly \emph{project} the input mesh $\mesh_I$ to the base mesh $\mesh_B$ by mapping each input vertex to its corresponding base location. For each point $p$ on the base, we then find the closest point on the projected mesh $\tilde{\mesh}_I$. This closest point gives us barycentric coordinates inside a projected mesh $\tilde{\mesh}_I$ triangle. We use the barycentric mapping between the projected and input triangles to obtain a corresponding position $\{p_S^0\}$ with these barycentric coordinates on the original mesh $\mesh_I$.

We note that this new position $\{p_S^0\}$ may not lie along the displacement direction defined at the point $p$.  We \emph{project} $\{P_S^0\}$ to the displacement ray at $p$ and get a new position $\{q_S^0\}$. After the projection, we find the closest point to $\{q_S^0\}$ on the input mesh $\mesh_I$, denoting as $\{p_S^1\}$. We then repeat the process, iteratively finding closest points on the mesh and ray. Because we keep finding closest points between $\mesh_I$ and ray, the distance between the mesh and ray points decreases at each iteration, ensuring that this iterative closest point algorithm will converge. Finally, we get $\{q_S^n\}$ as the position of the point $p$. As our results show, this is a more efficient and robust way to compute displacement value than ray-casting.

%% file: 5-results.tex
\section{Results}
\label{sec:results}
We evaluate our DJM method both qualitatively and quantitatively by comparing it to the micromesh construction method of Maggiordomo et al. ~\shortcite{Maggiordomo2023MicroMesh}, as well as QEM. We measure distance (RMS, max, and Chamfer) between the inputs and the reconstructed meshes.
We also show the use of our DJM base meshes for neural applications by encoding displacement mapping into a MLP. We evaluate our method for micromeshing on 89 shapes taken from Maggiordomo et al. ~\shortcite{Maggiordomo2023MicroMesh}'s dataset of 121 3D scans, consisting of 56 shapes that are originally watertight and manifold in the dataset, and a further 33 shapes with minor topological errors that can be manually fixed to be manifold and water-tight. The remaining shapes have over 20 open boundary or non-manifold edges, making them challenging to fix. For the 89 shapes, we detect degenerated triangles or ill-triangulated large planar regions (e.g. statue bases) and perform region-based remeshing \cite{hoppe1993mesh}. We evaluate our neural encoding on  20 shapes taken from the dataset of Zhang et al. ~\shortcite{zhang2024nesi}. We implement our algorithm in C++, using the PMP Library \cite{pmp23}, libigl \cite{libigl}, Eigen \cite{eigenweb}, and Maggiordomo et al's \cite{Maggiordomo2023MicroMesh} micromesh builder to compute adaptive subdivision levels. 

\subsection{Micromesh Comparison}
\begin{figure}[ht]
	\centering
	\includegraphics[width=0.85\linewidth]{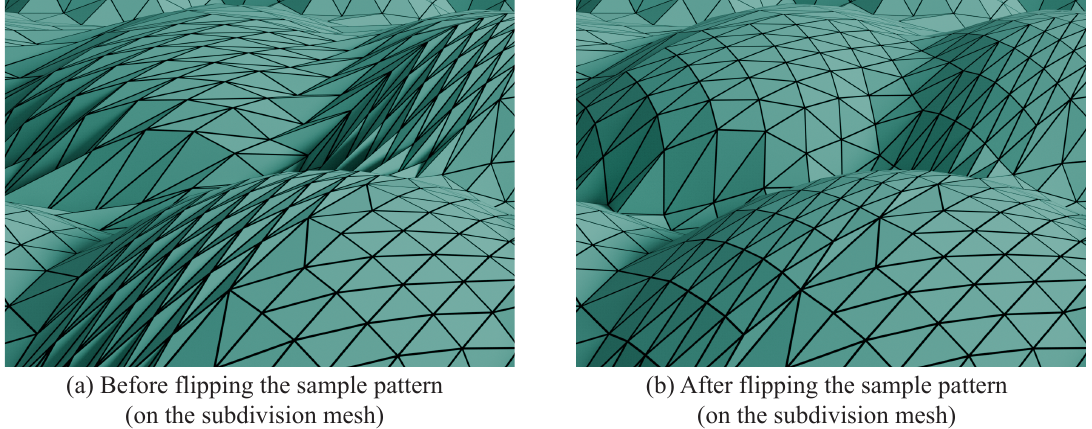}
	\caption{We optimize the sampling pattern on a per base-mesh triangle basis to alleviate Schwarz lantern artifacts.}
	\label{fig:schwarz}
\end{figure}
Fig. \ref{fig:mm_qem_comparison} shows a visual comparison between our DJM base meshes and displaced outputs, QEM base meshes and displacement mapped outputs, and those of Maggiordomo et al. ~\shortcite{Maggiordomo2023MicroMesh}. For each shape, we set the same target face count for the simplified base mesh, and set the subdivision target to 256 times the base mesh’s face count (which means the average subdivision level is 4). For all methods, the subdivision level of each face is determined by Maggiordomo et al.'s adaptive strategy. For both QEM and Maggiordomo et al., displacement maps are computed using the ray‑casting procedure provided in Maggiordomo et al.'s code. For DJM, we exploit our bijective mapping which allows us to compute displacement values for all vertices via our iterative projection algorithm (Sec. 4.1.2). To alleviate Schwarz lantern artifacts, we additionally check the average cosine similarity between the target surface normals and the subdivision face normals using our bijective mapping and flip the subdivision pattern if it can improve the normal similarity (Fig.\ref{fig:schwarz}).

We observe two major classes of artifacts in previous work that are resolved by our method. First, the base mesh is not always visible to the points they represent on the input surface, which leads to grazing regions (e.g. the hat in Fig. \ref{fig:mm_qem_comparison}, row 1; the goat's paws in Fig. \ref{fig:mm_qem_comparison}, row 4). Second, ray-casting is highly unstable when the base mesh is far from the target geometry, especially when the surface normal is nearly orthogonal to the casting direction. This results in spikes and failure cases (Fig. ~\ref{fig:ray_casting}, Sphinx model in Fig. \ref{fig:mm_qem_comparison}, row 7).

\begin{table}[ht]
	\tiny
	\centering
	\caption{Comparison of micromeshing metrics across different subdivision levels. File sizes are reported using the same encoding. (*For QEM, subdivision face counts/overall file size are lower due to poor mesh quality causing the raycasting subdivision algorithm to terminate early.)}
	\begin{tabular}{llccc}
		\toprule
		\textbf{Subdiv Level} & \textbf{Metric} & \textbf{Micro Mesh} & \textbf{QEM} & \textbf{DJM} \\
		\midrule
		\multirow{3}{*}{\textbf{256}} 
		& RMS (Mean/Median/Max) $\downarrow$ & 0.226/0.130/2.581 & 2.250/1.169/22.494 & \textbf{0.130/0.108/0.427} \\

		& Subdiv \#F (Median) $\downarrow$ & 421,380 & 327,956* & 365,754 \\
		& Out Size (Median) $\downarrow$ & 378,996 & 306,120* & 338,334 \\
		\midrule
		\multirow{3}{*}{\textbf{64}} 
		& RMS (Mean/Median/Max) $\downarrow$ & 0.408/0.284/3.091 & 3.921/2.277/36.479 & \textbf{0.320/0.252/1.168} \\

		& Subdiv \#F (Median) $\downarrow$ & 98,342 & 94,936 & \textbf{82,642} \\
		& Out Size (Median) $\downarrow$ & 133,878 & 126,344 & \textbf{123,756} \\
		\midrule
		\multirow{3}{*}{\textbf{16}} 
		& RMS (Mean/Median/Max) $\downarrow$ & 0.841/0.675/3.685 & 4.257/2.233/33.649 & \textbf{0.768/0.622/3.012} \\

		& Subdiv \#F (Median) $\downarrow$ & 23,138 & 28,552 & \textbf{21,268} \\
		& Out Size (Median) $\downarrow$ & 70,278 & 73,166 & \textbf{67,672} \\
		\bottomrule
	\end{tabular}
	\label{tab:mm_qem_comparison}
\end{table}
For the same base meshes, we evaluate the distance between the subdivision meshes and the original meshes at different target subdivision face counts: 256, 64, and 16 per base mesh triangle. The actual subdivision level is determined by Maggiordomo et al.~\shortcite{Maggiordomo2023MicroMesh}'s adaptive subdivision scheme, with a maximum subdivision depth of 5 (1024 faces) corresponding to hardware limitations. 
Tab. \ref{tab:mm_qem_comparison} evaluates geometric accuracy between ground truth meshes and displaced meshes generated by DJM and alternatives. We report the mean, median, and maximum RMS Hausdorff distances;  the median number of faces post subdivision; and the median output file size in bytes. Compared to prior art, DJM has improved geometric accuracy on all measurements. In particular, the significant lower RMS error of our method shows the robustness and stability of our DJMs. We provide full visual results along with per shape metric information in the supplementary material.

\paragraph{Runtimes.} Our method averages 15 minutes to compute the base mesh and 28 minutes to compute the displacement maps; see the supplemental for full details.

\subsection{Neural Displacement Comparison}
\label{sec:neural_results}
To evaluate the effectiveness of our base mesh for neural shape representation, we compare our method against NGF~\cite{Sivaram2024NGF} and the method of Pentapati et al. ~\shortcite{Pentapati2025MeshCompression}. Both our method and NGF use the same base‑mesh face count. NGF further groups triangles into quads and trains a neural vector field over this quad mesh. To ensure that our neural representation uses fewer total parameters, we assign NGF a 6-layer MLP and keep the configuration unchanged. Our model uses a 5-layer MLP, with all hidden layers 64 neurons; for our method, we train the neural network using 3 million pre‑sampled points uniformly distributed on the target surface, and compute the base mesh, directions, and displacements using our inverse solver. This enables direct uniform sampling on the target surface, which is infeasible with previous work. To ensure small base mesh triangles receive sufficient supervision, we sample an additional 3 million points, allocated evenly across all base mesh triangles. Each training iteration draws 50\% of its samples from uniform surface sampling and 50\% from equal‑triangle sampling.
Unlike NGF, which predicts a 3D vector field, our network predicts only a 1D displacement field defined on the base mesh. During training, we use an $L1$ displacement loss and a normal consistency loss. Our method outperforms NGF both visually and quantitatively on the NESI dataset \cite{zhang2024nesi}, which includes 20 manifold shapes.

\begin{table}[ht]
	\centering
	\scriptsize
	\setlength{\tabcolsep}{3pt}
	\caption{Quantitative comparison with NGF~\cite{Sivaram2024NGF}.}
	\begin{tabular}{lrrrrrr}
		
		\toprule
		Method & \#V & \#F & \#params & Chamfer $(10^{-3})$ & L2 $(10^{-5})$ & RMS $(10^{-3})$ \\
		\midrule
		NGF & 131 & 130 & 28944 & 0.609 & 1.53 & 0.766 \\
		Ours & 131 & 262 & \textbf{22247} & \textbf{0.252} & \textbf{0.481} & \textbf{0.285} \\
		\bottomrule
	\end{tabular}
	\label{tab:NGF_comparison}
\end{table}

We present the visual comparison in Figure~\ref{fig:NGF_comparison} and the quantitative results in Table~\ref{tab:NGF_comparison}. As both methods are evaluated on meshes, we generate extremely high resolution meshes for each approach to reduce any influence from meshing. As shown in both the visual and numerical results, our method yields cleaner, more faithful neural shape representations. This is partly due to the formulation of the displacement field: prior to training, the displacement at each sample point is already well defined, and the network simply overfits to these fixed pre-computed values. In contrast, NGF needs to optimize a full 3D vector field during training process, which is harder to supervise.

We also compare to Pentapati et al.~\shortcite{Pentapati2025MeshCompression} using their default 60k configuration. Over 13 shapes, their method's average Chamfer distance is 0.327; our method achieves an average Chamfer distance of 0.237 on the same shapes using fewer than 20k parameters (Fig. ~\ref{fig:SSP_comparison}). For further details see the supplemental.

\begin{figure*}[t]
	\centering
	\includegraphics[width=0.72\linewidth]{full_gallery_v3.pdf}
	\caption{Visual comparison of micromeshes constructed by Maggiordomo et al. ~\shortcite{Maggiordomo2023MicroMesh}, QEM~\cite{Garland1997QEM}, and our method.}
	\label{fig:mm_qem_comparison}
\end{figure*}

\begin{figure*}[t]
	\centering
	\includegraphics[width=0.72\linewidth]{NGF_gallery_v2.pdf}
	\captionof{figure}{Visual comparison with NGF~\cite{Sivaram2024NGF}.}
	\label{fig:NGF_comparison}
	\vspace{1em}
	\includegraphics[width=0.72\linewidth]{ssp_DJM_v3.pdf}
	\captionof{figure}{Visual comparison with Pentapati et al.\cite{Pentapati2025MeshCompression}.}
	\label{fig:SSP_comparison}
\end{figure*}

%% file: 6-conclusion.tex
\section{Conclusions}
\label{sec:conclude}
We present DJM, a new method for computing base meshes for displacement mapping, and demonstrate it to outperform the state-of-the-art alternatives in terms of accuracy to size trade-off for both micromesh construction and neural compression. The key elements that make this improvement possible are a new Jacobian-based error metric for evaluating base mesh construction operations; a novel progressive relaxation-based simplification strategy; and forward- and backward mapping methods that avoid the need for raycasting.

Future work includes optimization of base mesh vertex locations post construction, as well as additional strategies for avoiding and repairing isometric ``sliver'' triangles. It would also be interesting to explore additional neural compression applications beyond our basic solution that take advantage of our base meshes and correspondences.
\paragraph{Limitations.} Our current implementation only operates on closed meshes, following the default QEM implementation. Processing boundaries is conceptually straightforward, but requires careful implementation. Additionally, our bijectivity constraints may "bake-in" artifacts in the input mesh (non-manifoldness/foldovers/extreme variations in local normals). These artifacts can be mitigated by allowing the user to specify a threshold such that smaller non-bijective regions can be tolerated. Finally, we note that we only guarantee bijectivity in the sense that every collapse and flip operation during base mesh construction maintains a bijective mapping, similar to other works using successive self-parameterization (e.g. \cite{liu2020subdivision}). This is not necessarily a global guarantee of bijectivity due to numerical issues, but like other work it is sufficient for our purposes.

%% file: 7-acknowledgement.tex
\begin{acks}
We acknowledge the support of the Natural Sciences and Engineering Research Council of Canada (NSERC) grant RGPIN-2024-03981. Finally, this work is supported in part by the Institute for Computing, Information and Cognitive Systems (ICICS) and Advanced Research Computing (ARC) at UBC.
\end{acks}